\documentclass[fleqn,usenatbib]{mnras}

\usepackage{newtxtext,newtxmath}
\usepackage[T1]{fontenc}

\DeclareRobustCommand{\VAN}[3]{#2}
\let\VANthebibliography\thebibliography
\def\thebibliography{\DeclareRobustCommand{\VAN}[3]{##3}\VANthebibliography}

\usepackage{graphicx}	% Including figure files
\graphicspath{{./figures/}}
\usepackage{amsmath}	% Advanced maths commands
\usepackage{enumitem}
\usepackage{natbib}
\usepackage{hyperref}
\usepackage{xcolor}     % For highlighted notes; delete this when completed
\usepackage[normalem]{ulem}

\newcommand*\diff{\mathop{}\!\mathrm{d}}

\newcommand\kms{km~s$^{-1}$}
\newcommand{\pcm}{\,cm$^{-2}$}	% per cm-squared
\newcommand{\sub}[2]{#1_\mathrm{#2}}
\newcommand{\Lya}{Ly$\alpha$}

\newcommand{\Ncl}{\mathcal{N}_\mathrm{cl}}

\newcommand\nc{n_\mathrm{\scriptscriptstyle{H}}}
\newcommand\nN{n^{}_{{\scriptscriptstyle\mathcal{N}}_\mathrm{cl}}}
\newcommand\mcl{m_\mathrm{cl}}
\newcommand\betac{\beta_\mathrm{c}}
\newcommand\betaN{\beta^{}_{{\scriptscriptstyle\mathcal{N}}_\mathrm{cl}}}
\newcommand\normc{n_{\rm H,0}}
\newcommand\normN{n^{}_{{\scriptscriptstyle\mathcal{N}}_\mathrm{cl},0}}
\newcommand{\sigmac}{\sigma_{\nc}}
\newcommand\rcl{R_\mathrm{cl}}
\newcommand{\sigmacl}{\sigma_\mathrm{cl}}

\newcommand\rproj{d_\mathrm{proj}}

\newcommand{\vdisp}{v_\mathrm{disp}}

\newcommand{\sigmaleft}{\sigma_\mathrm{left}}
\newcommand{\sigmaright}{\sigma_\mathrm{right}}

\newcommand{\rvir}{\mbox{$r_{\rm vir}$}}

\newcommand{\msun}{\mbox{$\rm M_{\odot}$}}

\newcommand{\mstar}{\mbox{$M_{\rm star}$}}
\newcommand{\cc}{{\rm cm^{-3}}}

\title[Modeling Cool Clouds in CGM]{Modeling the cool gas clumps in the circumgalactic medium}

\author[H. Yang et al.]{
Hang Yang$^{1, 2, 3}$ \thanks{yanghang@pmo.ac.cn},
Zhijie Qu$^{4, 3, 5}$ \thanks{quzhijie@tsinghua.edu.cn},
Joel N. Bregman$^{3}$,
Li Ji$^{1, 6}$
\\
$^{1}$ Purple Mountain Observatory, Chinese Academy of Sciences, 10 Yuanhua Road, Nanjing 210023, China\\
$^{2}$ School of Astronomy and Space Sciences, University of Science and Technology of China, Hefei 230026, China\\
$^{3}$ Department of Astronomy, University of Michigan, Ann Arbor, MI 48109, USA \\
$^{4}$ Department of Astronomy $\&$ Astrophysics, The University of Chicago, Chicago, IL 60637, USA \\
$^{5}$ Department of Astronomy, Tsinghua University, Beĳing 100084, China \\
$^{6}$ Key Laboratory of Dark Matter and Space Astronomy, CAS, Nanjing 210023, China
}

\date{Accepted XXX. Received YYY; in original form ZZZ}

\pubyear{2025}

\begin{document}

\label{firstpage}
\pagerange{\pageref{firstpage}--\pageref{lastpage}}
\maketitle

\begin{abstract}
A major challenge in CGM studies is determining the three-dimensional (3-D) properties from the observed projected observations.
Here, we decompose the 3-D gas density and spatial distribution of cool clouds by fitting a cool CGM model with the absorption observations, including the cool gas density, Ly$\alpha$, and \ion{Mg}{II} equivalent widths.
The clumpiness in the cool CGM is considered by modeling individual clouds.
This model has four major components: the radial profile of the cool gas density; the number density of clouds; the absorption properties within individual clouds; and the velocity dispersion in the CGM.
%The model is constrained in a two-step fitting.
The observed cool gas density exhibits a large dispersion of $\approx 2-3$ dex within the virial radius ($\rvir$).
This dispersion can be reproduced with a combination of the projection effect (i.e., distant low-density clouds projected at small radii) and the intrinsic variation in the gas density.
By modeling the probability density functions of gas density at different radii, the cool gas density is modeled as a $\beta$-model with a characteristic gas density of $\log n_{\rm H,0}/\cc =-2.57_{-0.25}^{+0.43}$ at $\rvir$ and a slope of $\beta_c = 0.63_{-0.20}^{+0.16}$, and the intrinsic dispersion is $\sigma_{\nc}\approx 0.56_{-0.20}^{+0.19}$ dex.
Assuming a cloud mass of $10^4~\msun$, we further constrain the number density of cool clouds by jointly reproducing Ly$\alpha$ and \ion{Mg}{II} equivalent width samples, resulting into a number density of $\log \normN/ \rvir^{-3} = 4.76^{+0.27}_{-0.21}$ at $\rvir$ and a slope of $\betaN = 0.65^{+0.06}_{-0.07}$.
This spatial distribution of the cool CGM leads to a total cool gas mass of $\log M_{\rm cool}/\msun = 10.01^{+0.06}_{-0.06}$ for $L^*$ galaxies, while varying the cloud mass from $10^3~\msun$ to $10^6~\msun$ leads to the total cool CGM mass of $9.62_{-0.07}^{+0.05}$ to $10.46_{-0.05}^{+0.05}$.
%This cool CGM modeling framework can be extended with more ions and emissions in future works.
\end{abstract}


%%%%%%%%%%%%%%%%%%%%%%%%%%%%%%%%%%%%%%%%%%%%%%%%%%

%%%%%%%%%%%%%%%%% BODY OF PAPER %%%%%%%%%%%%%%%%%%
\section{Introduction}
Galaxies are surrounded by massive gaseous atmospheres, known as the circumgalactic medium (CGM).
This gaseous component is crucial in understanding the baryonic cycle of galaxy evolution, maintaining continuous accretion to galaxies for star formation and gathering feedback from stellar or AGN activities \citep[see reviews][]{2022PhR, 2023ARAA}.
In the past three decades, extensive multi-wavelength observations reveal the multiphase nature of the CGM, with density and temperature spanning over decades, including cool phase ($T \sim 10^4$~K) gas, warm phase ($\sim 10^5$~K) gas, and hot phase gas ($\gtrsim 10^6$~K; e.g., \citealt{2017ARAA, Chen2024}).
However, a fundamental question is how different gas phases are distributed in galaxy halos. 

QSO spectroscopy has been a powerful tool for detecting and measuring the diffuse gas in the CGM, especially for the cool CGM over the last billion years
\citep[e.g.,][]{Chen1998, Chen2001, Steidel2010, Bordoloi2014, Werk2014, Johnson2015, Johnson2017, Rudie2019, Lehner2020, Wilde2023, qu2023MNRAS}.
By resolving the absorption features of specific ions, insights have been obtained in the spatial distribution and kinematics of the cool CGM,  \citep[e.g.,][]{Martin2019, Ho2019, dutta2020MNRAS, huang2021MNRAS}.
However, serious problems still exist in converting the observed properties into physical properties for two major reasons.
First, the cool CGM is clumpy, leading to significant dispersion in observed properties.
Second, the observed properties are integrated over the path length, making it challenging to reconstruct the 3-dimensional (3D) physical distribution or kinematics.

Various models have been established to infer cool gas physical properties from absorption data. 
Typically, the observed absorption strengths are produced with assumed density or pressure profiles, together with cool gas conditions, such as under the photoionization equilibrium 
\citep[e.g.,][]{Stern2016, faerman2023ApJ, hummels2023arXiv, Faerman2024, Faerman2025, li2024MNRAS}.
Because of the complicated structures and clumpiness of the cool CGM, more and more recent works consider the density variations in the cool CGM \citep[e.g.,][]{Stern2016, DuttaA2024, Faerman2024, Bisht2024}.

In observations, recent high-signal-to-noise and high-spectral-resolution UV and optical spectroscopy make it possible to resolve small-scale cool gas structures, constraining the density, temperature, and metallicity for individual clouds in the cool CGM \citep[e.g.,][]{Cooper2021, Haislmaier2021, Sameer2021, Sameer2024, zahedy2019MNRAS, Zahedy2021, qu2022MNRAS, Kumar2024}.
The observed cool gas densities of different absorption components can vary over three decades in one sight line, while the origin of this scatter is still uncertain.
In principle, the density scatter exhibits two origins.
First, the density scatter may be due to the projection effect.
The global density or pressure profile is declining to the outskirts, and distant low-density gas may be projected close to the galaxy.
This leads to large density scattering at a small projected distance.
Another possibility is the intrinsic density and pressure variations in the cool gas, induced by rapid cooling \citep[e.g.,][]{Stern2020} or potential non-thermal processes (e.g., cosmic ray; \citealt{Butsky2020, Ji2020, CR2023AARv}).

Investigating the origin of density scatter is still an open question, which needs to incorporate the clumpiness in the cool CGM modelling.
Here, we introduce a framework considering the cool CGM clumpiness and model the observed probability distribution function (PDF) of observed gas densities and equivalent widths ({\rm EW}s) at different projected distances.
This paper is organized as follows.
In Section \ref{sec:model}, we introduce the adopted assumptions and the model setup.
Then, Section \ref{sec:data} summarises the data used in this study.
The results are presented in Section \ref{sec:analysis} after fitting the model to the observations.
In Section \ref{sec:dis}, we discuss the caveats in the model and the model implications.
Section \ref{sec:sum} summarizes the key findings.
Throughout the paper, we adopt a $\Lambda$ cosmology with $\Omega_{\rm m} = 0.3$, $\Omega_{\Lambda} = 0.7$, and a Hubble constant of $H_0 = 70\rm ~ km ~s^{-1}~ Mpc^{-1}$.

\section{The Cool Cloud CGM Model}
\label{sec:model}

In this section, we introduce the basic assumptions in the cool cloud CGM model, which involves both intrinsic variation and projection effect to reproduce the observed absorption properties (i.e., gas density and equivalent width).
In particular, there are four components in the model: (i) the global radial profile of the cool phase gas density and the associated local density scatter; (ii) the radial profile of the number density of cool clouds; (iii) the ionization modeling of individual clouds; and (iv) the broadening of absorption features.
Then, the PDFs along sightlines projected at different distances are predicted for the gas density, ionic column density, and corresponding equivalent width.
This model will be fitted to observations in the following Sections \ref{sec:data} and \ref{sec:analysis} to obtain the best fit of the cool gas distribution.

\subsection{Basic assumptions}
The cool clouds in the CGM typically have sizes of $\sim 0.1$ kpc \citep[e.g.,][]{Lehner2019ApJ, Zahedy2021, Sameer2024}.
Therefore, we consider a clumpy gas model, where the cool CGM consists of numerous discrete small clouds embedded in the hot halo, which is also consistent with simulation predictions \citep[e.g.,][]{Lizhihui2020, Ramesh2023}.
For simplicity, we assume cool clouds are spherically and isotropically distributed in galaxy halos.
Then, the cool gas density, physical size, and number density of cool clouds are represented by one-dimensional profiles.

\subsubsection{Cool gas density profile }
\label{sec:density_profile}

The cool gas is embedded in the hot CGM as individual clouds.
In observation, individual clouds can be resolved by decomposing absorption components based on line centroids (but also see \citealt{hummels2023arXiv} for potential mixing in individual components).
Within each cloud, the gas density is roughly constant, although in some cases, multiple phases are needed to explain the observed column density ratios between metal transitions for one absorption component \citep[e.g.,][]{zahedy2019MNRAS, qu2022MNRAS}.
However, the derived cool gas density varies depending on how component decompositions and ionization mechanisms are treated \citep[e.g.,][]{Werk2014, Lehner2019ApJ, Sameer2024}.
These factors lead to different cool gas density profiles.
Despite uncertainties in the slope and amplitude, a power law profile is often used to model the observed density.

In this study, we adopt the $\beta$-model to describe the gas density profile $\nc(r)$, which is often adopted to model the hot gas distribution in the halo \citep[e.g.,][]{bogdan2013ApJ, miller2013ApJ, li2018ApJ}:
\begin{equation}
      \label{eq:nc}
    \nc(r) = \frac{n_{\rm H,0}}{A} \left[1 + \left(\frac{r}{r_\mathrm{core}}\right)^2\right]^{-\frac{3}{2} \betac},
\end{equation}
where $n_{\rm H, 0}/A$ is the normalization of the cool gas density profile and $\betac$ is the slope beyond the core radius ($r_\mathrm{core}$).
Here,  $\normc$ is the gas density at the virial radius ($\rvir$) with a normalization term of 
\begin{equation}
    A = \left[1 + \left(\frac{r_{\rm vir}}{r_\mathrm{core}}\right)^2\right]^{-\frac{3}{2} \betac}.
\end{equation}
The core radius $r_\mathrm{core}$ is fixed to $ 0.01~ \rvir$ to avoid singularity in the center.
Such a core radius is significantly smaller than the minimum probed distance, which is insensitive to other parameters.

\subsubsection{Cool cloud mass}
\label{sec:assump:mcl}
To consider the clumpiness of the cool CGM, the cool gas is assumed to be spherical clouds for simplicity.
Once the gas density is obtained from the density profile, we adopt a cloud mass of $\mcl= 4\pi\rcl^3 \mu \nc m_{\rm p}/3$ ($\mu=1.4$ accounting for helium) to calculate the cloud size ($\rcl$) and column densities for individual clouds in the following modelling (Section \ref{sec:prob:Nion}).
Driven by the anti-correlation found between the inferred density $\sub{n}{H}$ and clump size $\sub{R}{cl}$ \citep{chen2023ApJ}, we assume an empirical and constant cloud mass of $10^4~ \msun$ in the model.
Different choices of cloud masses will be further discussed in Section \ref{sec:discuss:mcl}.

\subsubsection{Number density profile of clouds}
\label{sec:Ncl_profile}
After obtaining the properties of individual clouds, the spatial distribution of cool clouds is needed to calculate the integrated column density and observed equivalent width.
Although cool clouds may have large-scale coherence in the halo \citep{Rubin2018}, we do not consider this effect in this work, because of few observational constraints.
For simplicity, we assume an isotropic radial profile of the cool clouds in this study, from which we will extract a discrete number of clouds under the Poisson assumption after considering the projection effect (described in Section \ref{sec:density_dis}).

It remains unclear how these clouds are distributed in the halo.
Here, we adopt a modified $\beta$-model, with a truncation at a large radius.
The number of clouds per unit space volume is
    \begin{equation}
      \label{eq:nN}
      \nN (r)= \frac{n_{{\scriptscriptstyle\mathcal{N}}_\mathrm{cl},0}}{B}
      \left[1 + \left(\frac{r}{r_\mathrm{core}}\right)^2\right]^{-\frac{3}{2} \betaN} 
      \exp \left[-\left(\frac{r}{r_\mathrm{halo}}\right)^{\beta_\mathrm{halo}}\right],
  \end{equation}
where $\betaN$ is the slope of the number density profile, and $r_\mathrm{halo}$ is the cutoff radius for the gaseous halo, beyond which IGM or the second-halo term will dominate the absorption features.
In particular, we fix $r_\mathrm{halo} = 2\rvir$, a typical boundary between the primary halo and second halo in \ion{H}{I} clustering analysis \citep[e.g.,][]{Morris1993, Chen2009, Wilde2023}.
The parameter $\beta_{\rm halo}$ is fixed to 2, setting a smooth boundary of the halo, which is insensitive to the model predictions.
Similar to the gas density profile,  $n_{{\scriptscriptstyle\mathcal{N}}_\mathrm{cl},0}$ is obtained at the virial radius with an additional normalization term of
\begin{equation}
    B = \left[1 + \left(\frac{\rvir}{r_\mathrm{core}}\right)^2\right]^{-\frac{3}{2} \betaN} 
      \exp \left[-\left(\frac{\rvir}{r_\mathrm{halo}}\right)^{\beta_\mathrm{halo}}\right].
\end{equation}

\begin{figure*}
  \includegraphics[width=0.97\textwidth]{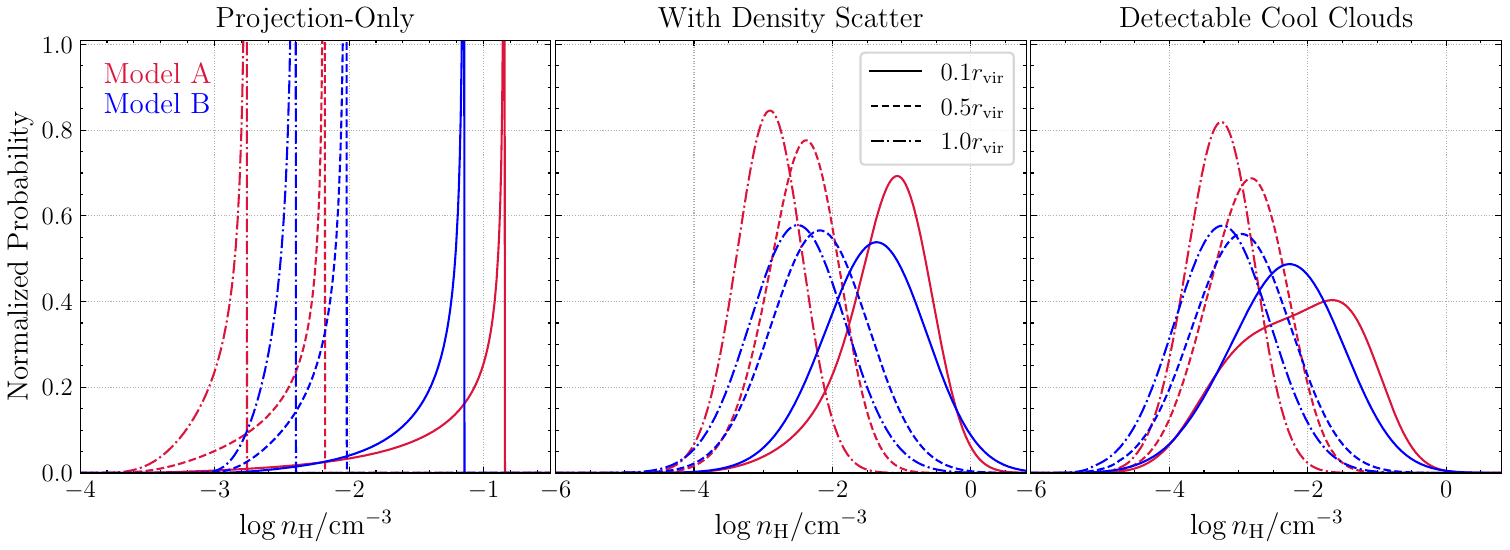}
  \caption{The model-predicted PDFs of the gas density. Two models $\nc$ (A) and (B) are shown with different contributions of the projection effect and the intrinsic dispersion, where model $\nc$ (A) is more contributed by the projection effect, while (B) is dominated by the intrinsic scatter (detailed in Table \ref{tab:best_fit} and Section \ref{sec:nH_fit}).
  Left panel: the projection-only model calculated using Equation \ref{eq:prob_logn}. The maximum density is calculated from Equation \ref{eq:nc}. Middle panel: the model prediction including the density scatter.
  The PDF width of Model $\nc$ (A) exhibits a significant decrease with the radius, while Model $\nc$ (B) shows similar widths. 
  Right panel: The PDF with the cross-section correction represents the case when clouds are detectable. This correction leads to the PDF having more low-density clouds.
  }
  \label{fig:nH_comparison}
\end{figure*}

\subsection{Intrinsic variation and projection effect in the gas density}
\label{sec:density_dis}

Recent observations reveal significant variations in the cool gas density, which may have different implications for the physical conditions in the cool CGM.
First, the cool gas density is not the same at a given 3D radius, leading to the intrinsic scatter of the density.
Therefore, the observed density scatter can be due to the halo-by-halo difference or the line-of-sight difference in a given galaxy halo, because there is normally only one sightline for each host galaxy.
Second, a significant decline over the radius of the cool gas density is observed \citep[e.g.,][]{zahedy2019MNRAS, qu2023MNRAS, Sameer2024}, implying that the distant low-density clouds may be projected at a small distance.
In this model, we consider both contributors to the observed density scatter. %, and decompose them empirically.

First, we adopt an intrinsic scatter in the logarithm scale ($\sigma_{\nc}$) for the gas density.
This scatter factor is applied to the gas density profile defined in Section \ref{sec:density_profile}, which leads to a log-normal probability distribution at $r$, instead of a $\delta$ function without the intrinsic scatter.
This factor is assumed to be a constant over different radii, because it is mostly unconstrained in observations or simulations \citep[e.g.,][]{Ji2020}.
However, we note that this assumption may not be practical.

Then, we derive the probability of observed $\nc$ at each projected distance ($\rproj$) for the projection effect by combining the cool gas density profile and the number density profile of individual clouds.
We start with the case without the density scatter, where each density corresponds to a specific 3D radius  $r(\nc)$ following the reverse function of Equation \ref{eq:nc}:
\begin{equation}
\label{eq:prob_n}
\begin{aligned}
  \rho (\nc | \rproj) &= \rho(r | \nc, \rproj) \delta(r = r(\nc))\\
  &= \sigmacl  \nN  \left[\frac{r(\nc)}{\sqrt{r(\nc)^2 - \rproj^2}} \frac{{\rm d} r}{{\rm d} \nc}\right].
\end{aligned}
\end{equation}
In this equation, $\rho(r | \nc, \rproj)$ is the probability of different radii to have $\nc$ at a given $\rproj$, while $\delta (r=r(\nc))$ ensures only one radius contributes to one observed density.
The cross-section of cloud $\sigmacl = \pi \rcl^2$ has a dependence on the gas density under the constant cloud mass assumption, then $\sigmacl  \nN$ is the number density of clouds along the path length.
The last term corrects the geometry to convert the path length along the radial direction to the line of sight at $\rproj$.

To combine the intrinsic scatter and the projection effect, we modify Equation \ref{eq:prob_n} with a two-step method.
First, the $\nc$ probability distribution is converted into the logarithm scale by 
\begin{equation}
\label{eq:prob_logn}
  \rho(\log \nc | \rproj) = \rho(\nc | \rproj) \cdot \nc / \log_{10} e.
\end{equation}
Then, instead of a unique density at a given $r$, clouds may exhibit different densities following a Gaussian function with a scatter $\sigma_{\nc}$ in the logarithm scale.
The final density distribution at $\rproj$ can be obtained by convolving the zero-mean Gaussian of  $G(0, \sigma_{\nc})$ with the density distribution without the intrinsic scatter.
However, we note that the Gaussian function cannot be convolved with Equation \ref{eq:prob_n} directly.
The detection probability depends on the cross-section ($\sigmacl$), which is relevant to the gas density rather than the cloud location.
Therefore, $\sigmacl$ is applied in the calculation after the convolution is done to account for the potential density scatter at a given radius:
\begin{equation}
\label{eq:prob_n_sigma}
\begin{aligned}
  \rho (\log \nc | \rproj, \sigma_{\nc} ) = \left[\frac{\rho(\log \nc | \rproj)}{\sigmacl} * G(0, \sigma_{\nc}) \right] \sigmacl .
\end{aligned}
\end{equation}
In the following analysis, $\sigma_{\nc}$ is always considered, so we omit it in all relevant equations.

With Equation \ref{eq:prob_n_sigma}, we extract two other properties for the following modeling and analyses.
First is the expected number of clouds ($\mathcal{N}_{\rm cl}$) along the line of sight at $\rproj$, which is a direct integral of Equation \ref{eq:prob_n_sigma} overall different densities.
The expected number of clouds will be used to determine the detection rate of absorbers in observations.
Another one is the normalized probability of the gas density $p (\log \nc | \rproj)$\footnote{We use $\rho$ for detection or incidence rates of clouds with specific $\log n_{\rm H}$ or ion column densities in the following sections, and $p$ for corresponding normalized probability distributions.} at $\rproj$ if any clouds are hit along the line of sight, which is directly relevant to the probability distribution of absorption properties.

In Figure \ref{fig:nH_comparison}, we show the effect due to projection and intrinsic scatter in the gas density PDF for two models $\nc$ (A) and (B), which are defined in Table \ref{tab:best_fit}.
These two models are two potential solutions with different relative contributions of the projection effect and the intrinsic scatter, which will be obtained in Section \ref{sec:nH_fit}.
The left panel in Figure \ref{fig:nH_comparison} shows the projection-only PDF of gas density at different radii of $0.1$, $0.5$, and $1.0~\rvir$, while the middle and right panels include the density scatter and the cross-section correction, respectively.
Here, we note that the median of the $\nc$ PDF is about $\approx 0.5-1.5$ dex lower than the gas density calculated using Equation \ref{eq:nc}.

\subsection{Probability distribution of column density}
\label{sec:prob:Nion}

For the observed ionic column density profiles, the scatter can be more than a factor of $10$ at a given projected distance, no matter whether it is normalized by $\rvir$ or not \citep[e.g.,][]{Chen2001, werk2013ApJS, qu2023MNRAS, Mishra2024}.
To capture the scatter in the column density, we adopt a two-step calculation.

First, we consider the probability distribution of ionic column density in individual clouds, which is a direct consequence of the probability distribution of the gas density (Equation \ref{eq:prob_n_sigma}).
As stated in \ref{sec:assump:mcl}, we adopt a constant cloud mass in the fiducial model, so different gas densities lead to different total hydrogen column densities per cloud ($N_{\rm H}$).
Then, we obtain $N_{\rm H} = \mcl/\pi\mu \rcl^2$, which is constant over the cloud cross section, ignoring the subtle structures within individual clouds.

Then, for each $\nc$ and its corresponding $\sub{N}{H}$, ion fractions are obtained from the photoionization equilibrium (PIE) model using the \textit{Cloudy} v17 \citep{Ferland2017}.
The cool CGM can be well modeled under PIE with a cosmic ultraviolet background (UVB; e.g., \citealt{Werk2014, qu2022MNRAS, Kumar2024}).
Here, we adopt the HM05 UVB model (an updated \citealt{Haardt2001} UVB in Cloudy), because the gas density sample in this study is obtained using the HM05 UVB (also see recent UVB; e.g., \citealt{Khaire2019, FG20}).
According to \citealt{zahedy2019MNRAS}, the major difference is that the derived density using the HM05 UVB is higher than FG20 by 0.2 dex, while other parameters (e.g., metallicity) show little systematic differences ($<0.1$ dex).
Therefore, using a different UVB model could result in a different gas density normalization ($n_{\rm H,0}$), but will not affect the derived slopes, total hydrogen column density, and cool gas mass.
The impact of the incident radiation field will be further discussed in Section \ref{sec:caveats}.

\begin{figure}
    \centering
    \includegraphics[width=0.97\columnwidth]{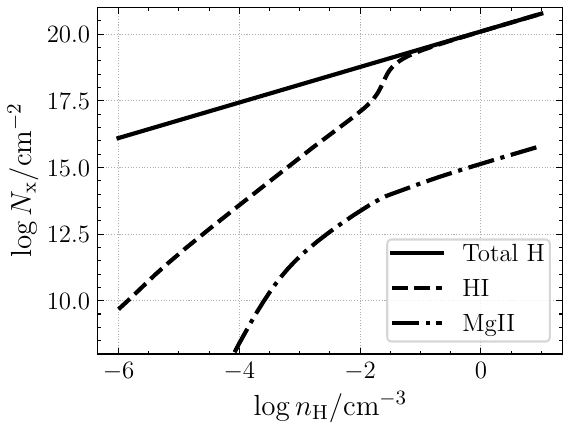}
    \caption{The fiducial Cloudy model of the total hydrogen ($N_{\rm H}$), \ion{H}{I} and \ion{Mg}{II} column densities under the photoionization under the HM05 UVB. Here, clouds with different densities exhibit constant mass as assumed in Section \ref{sec:assump:mcl}.}
    \label{fig:Nion_interp}
\end{figure}

In particular, we run a grid of models with $\log \nc/{\rm cm}^{-3}$ ranging from -6 to 1 with 0.1 dex steps, and $\log \sub{N}{H}/{\rm cm}^{-2}$ is calculated with the fixed cloud mass (see Section \ref{sec:assump:mcl}).
In addition, a finer step size of 0.01 dex is adopted, where $\sub{N}{\rm HI}$ increases dramatically at $\nc = 10^{-2}$ to $10^{-1}~\cc$.
In these models, the metallicity is fixed to $0.3~Z_\odot$, an average in the CGM \citep{prochaska2017ApJ, Zahedy2021}.
Although metallicity and abundance patterns can vary significantly in observations, their variation is beyond the scope of this study.
Also, a temperature floor is set at $8 \times 10^3$~K, which meets the PIE condition.
Thus, we obtain a tabulated model to map the gas density to different ion column densities under the PIE assumption.
Total hydrogen column density ($N_{\rm H}$), and \ion{H}{I} and \ion{Mg}{II} column densities as functions of the gas density are shown in Figure \ref{fig:Nion_interp}.

With the Cloudy models, we convert the PDF of gas density $p (\log \nc | \rproj)$ to the PDF of column densities for target ions $p(\log N_{\rm ion, cl}|\rproj)$, where ``cl'' denotes individual clouds.
Then, the PDF of the integrated column density is calculated by combining $p(\log N_{\rm ion, cl}|\rproj)$ and the expected cloud numbers $\mathcal{N}_{\rm cl}$.
In particular, different clouds are assumed to be independent of each other in Section \ref{sec:Ncl_profile}.
Here, we consider different cases if the sight line hits different numbers of clouds ($k$) following the Poisson distribution.
First, the probability of detecting $k$ clouds from a Poisson distribution with $\Ncl$ is ${\rm Pois}(k|\Ncl)$.
Then, we calculate the summed probability distribution  $p_k(\log N_{\rm ion}|\rproj)$ for $k$ clouds with independent $p(\log N_{\rm ion, cl}|\rproj)$.
Finally, we convolve the $p_k(\log N_{\rm ion} | \rproj)$ with its Poisson weights to generate the probability distribution of total ion column density at different projected radii:
\begin{equation}
  p(\log N_{\rm ion}|\rproj) =
  \left\{\begin{aligned} &
  \sum_{k=1} {\rm Pois}(k|\Ncl) p_k(\log N_{\rm ion}|\rproj), ~ k \geq 1 \\
  &{\rm Pois}(0|\Ncl) \delta(N=0), ~k=0.
\end{aligned}\right.
\end{equation}
The PDF of the total column density exhibits two different cases, depending on whether a cloud is hit along the sightline.
When there is no cloud along the sightline, the ion column density is zero in the model, whose probability is only determined by the Poisson distribution.
In the following analyses, we define the incidence rate of any clouds $\epsilon \equiv 1-{\rm Pois}(0|\Ncl)$, irrespective of whether clouds can be detected or not.
For any detected absorbers, the probability distribution is the sum of all cases with different numbers of clouds along the sightline.

\begin{figure*}
  \includegraphics[width=0.97\textwidth]{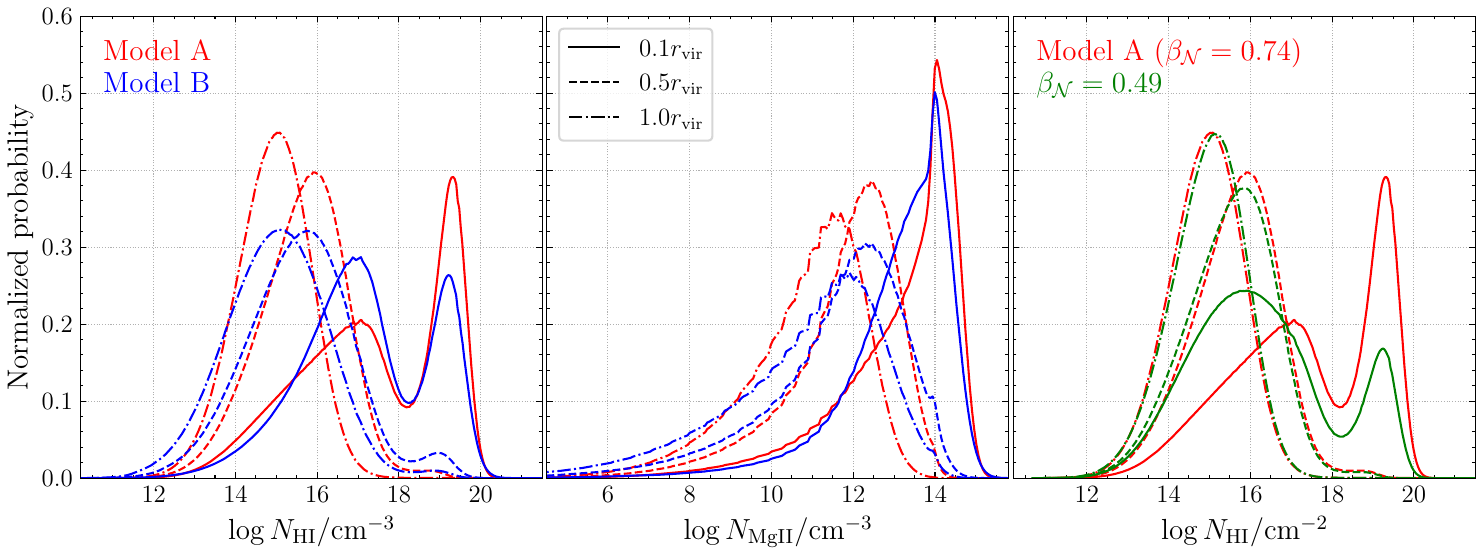}
  \caption{The model-predicted PDFs of the column density. Left and middle panels: the probability distribution of the \ion{H}{I} and \ion{Mg}{II} column density for Models $\nc$ (A) and (B).
  The double-peak in the \ion{H}{I} column distribution is a result of the sharp increase at $\log N_{\rm HI}\approx 17-18$ due to self-shielding (see Figure \ref{fig:Nion_interp}).
  Right panel: the comparison of $\log N_{\rm HI}$ distributions between two models with different slopes of the number density of clouds.
  Here, the total numbers of cool clouds within the virial radius are the same between the compared models.
  In the shallower slope case, the PDFs of column density exhibit similar median column densities at different projected distances.
  }
  \label{fig:Nion_comparison}
\end{figure*}

Similar to Figure \ref{fig:nH_comparison}, we show the model-predicted PDFs of column density PDFs in different models in Figure \ref{fig:Nion_comparison}.
In the left and middle panels, we show the PDF of \ion{H}{I} and \ion{Mg}{II} column densities, respectively.
Specifically for \ion{H}{I}, the double-peak distribution is a result of the sharp increase at $\log N_{\rm HI}\approx 17-18$ due to self-shielding. 
The right panel of Figure \ref{fig:Nion_comparison} shows a comparison between two PDFs of models with different slopes of the cloud number density profile.
For the model with a shallower slope of $\betaN=0.49$, the PDFs of column density exhibit similar median column densities at different projected distances.

\subsection{Probability distribution of equivalent width}
\label{sec:prob:EW}
After obtaining the PDF of the integrated column densities, we can further calculate the probability of the EWs for target ions, which are measured in spectroscopy observations \citep[e.g.,][]{Borthakur2015, Johnson2015, Johnson2017, Bordoloi2018, huang2021MNRAS}.
To convert column densities to EWs, the velocity dispersion of absorption features is required.
Decomposing absorption features into individual Voigt components in high-resolution QSO spectroscopy, the line width (i.e., Doppler $b$ factor) of \ion{H}{I} components is typically $\approx 30$ \kms \citep[e.g.,][]{Rauch1998}, while metal lines have smaller Doppler $b$ factors of $\approx 10$ \kms \citep[e.g.,][]{Rauch1996, Zahedy2021}.
However, multiple absorption components are detected along individual sightlines probing the CGM \citep[e.g.,][]{werk2013ApJS, Sameer2021}.
Then, the observed EW is more dominated by the velocity dispersion over all different absorption components ($\vdisp$), which can be as large as $\approx 100$ \kms in the CGM \citep[e.g.,][]{Koplitz2023, Qu2024}.

However, the PDF of velocity dispersion remains uncertain in observations.
Here, we adopt the split Gaussian distribution with asymmetric widths to approximate the velocity dispersion distribution:
\begin{equation}
\label{eq:b_scat}
    p(\vdisp) = 
    \begin{cases}
    \sub{A}{left} G(\vdisp |\sub{b}{peak}, \sigma_{\mathrm{left}}), &\, \vdisp\leq\sub{b}{peak}\\
    \sub{A}{right} G(\vdisp |\sub{b}{peak}, \sigma_{\mathrm{right}}), &\, \vdisp>\sub{b}{peak}\\
    \end{cases}
\end{equation}
where $\sub{A}{left}$ and $\sub{A}{right}$ are the normalization factors to make sure the Gaussian functions on either side are continuous at the peak position.
The free parameters $\sub{b}{peak}$, $\sigmaleft$, and $\sigmaright$ are the peak, the left width, and the right width of the split Gaussian distribution, which will be determined empirically with adopted data in Section \ref{sec:joint_lya_mgII}.

With the $\vdisp$ and $\log N_{\rm ion}$ PDFs, we calculate the EW PDF.
At a given $\vdisp$, EW and $N$ are monotonically correlated, and the corresponding distribution of EW can be obtained by following the conservation of probability:
\begin{equation}
  p(\log {\rm EW}|\rproj, \vdisp) \diff \log {\rm EW} = p(\log N|\rproj) \diff \log N,
\end{equation}
where the EW values are calculated from the column densities using the curve of growth for a given $\vdisp$.
Then, we convolve the EW distribution with the $\vdisp$ probability distribution (Equation \ref{eq:b_scat}):
\begin{equation}
\label{eq:prob:EW}
     p(\log {\rm EW} | \rproj) = 
     \left\{\begin{aligned} & \int p(\log {\rm EW}| \rproj, \vdisp) p(\vdisp) {\rm d} \vdisp, k\geq 1,\\
     &{\rm Pois}(0|\Ncl) \delta({\rm EW}=0), ~k=0.
     \end{aligned}\right.
\end{equation}
Similar to the PDF of column density, $p (\log {\rm EW}|\rproj)$ also exhibits two cases, depending on whether a cloud is hit along the sightline.
Here, the modeled EW has the dependence on parameters of $n_{\rm H,0}$, $\betac$, $\mcl$, $n_{{\scriptscriptstyle\mathcal{N}}_\mathrm{cl},0}$, $\betaN$, $v_{\rm peak}$, $\sigma_{\rm left}$, and $\sigma_{\rm right}$, and the UVB model.
This equation is adopted to compare with the observed EW distribution in the following analysis.

\section{Data}
\label{sec:data}
The cool cloud CGM model is constrained using multiple QSO absorption observations.
Ideally, the density profile and number density profile of clouds can be simultaneously constrained using the absorption strength and relative ratios of ionic transitions in galaxy surveys with QSO sightlines probing the CGM. 
However, considering the complicated structures in multiscale and multiphase CGM, we adopt a two-step fitting method to simplify the analyses in this study.
First, we adopt a sample of cool gas densities in the CGM directly derived from detailed photoionization modeling for decomposed absorption features.
This gas density sample is included to mainly constrain the gas density profile and the intrinsic density scatter.
Then, absorption strengths (i.e., EW) of Ly$\alpha$ and \ion{Mg}{II} are adopted to constrain the number density profile of cool gas clouds in the CGM.
In this section, we summarize the observations adopted in this study.

\subsection{The cool gas density sample}
The cool gas density can be derived with assumed ionization mechanisms \citep[e.g.,][]{Werk2014, Lehner2019ApJ}.
To break the degeneracy in gas density dispersion between the projection effect and the intrinsic scatter, the probability distributions of gas density at different projected distances are needed.
Therefore, we adopt studies with densities obtained for individual components decomposed using Voigt profile fitting, including the COS-LRG survey \citep{zahedy2019MNRAS} and several studies from the Cosmic Ultraviolet Baryon Survey (CUBS) program\footnote{There are also similar analyses, but adopted different assumptions in the ionization models \citep[e.g.,][]{Haislmaier2021, Sameer2021, Sameer2024}. In addition, these studies exhibit different observation depths for nearby galaxies, affecting the association between QSO sight lines and galaxies. In \citet{qu2023MNRAS}, all absorbers in one along are assigned to the galaxy with the smallest $\rproj/\rvir$ in deep galaxy surveys, which typically have a limiting stellar mass $\log \mstar/\msun \approx 8$ up to $z \approx 1$. To minimize systematical differences between different studies, we did not include these studies in the analyses.} \citep[i.e.,][]{Cooper2021, Zahedy2021, qu2022MNRAS}.

The gas density sample adopted in this study is compiled in \citet{qu2023MNRAS}\footnote{In \citet{qu2023MNRAS}, the halo size is characterized as $r_{200}$, within which the average dark matter density is 200 times the cosmic critical density. Here, we adopt the virial radius, where the overdensity is calculated using the relation in \citet{Bryan1998}. Typically, $\rvir$ is about 20-30\% larger than $r_{200}$.}.
Totally, there are 26 galaxies with stellar masses ranging from $\log \mstar/\msun \approx 8-11.5$ at $0.3 \lesssim z\lesssim 1.0$.
In particular, the full sample is about evenly split between star-forming and quiescent galaxies.
Limited by the currently available sample, we assume that different galaxies share similar density profiles, which cannot be distinguished statistically.

The gas density spans over three decades from $\log n_{\rm H}/{\rm cm^{-3}} \approx -4$ to $-1$, showing a decline from the inner halo to the outskirts.
Using the same gas density sample, \citet{qu2023MNRAS} report a power law between the maximum densities along individual sightlines and projected distance as $\log (n_{\rm H}/\cc) = (-1.9\pm 0.3) \times \log (d_{\rm proj}/r_{\rm vir})] + (-3.0\pm0.1)$ with an intrinsic scatter of $\approx 0.4$ dex.
This relation assumes that the maximum density along the sightlines also has the smallest 3D distance (i.e., equal to the projected distance).
With the cool cloud model, we will test whether this density decline is consistent with the low-density clouds in sightlines after considering the projection effect (Section \ref{sec:nH_fit}). 

\subsection{The EW sample of \ion{H}{I} Ly$\alpha$ and \ion{Mg}{II} $\lambda 2796$}
After obtaining the gas density profile, the absorption strength is needed to constrain the spatial distribution of cool clouds.
In this work, we adopt two typical transitions with large samples, Ly$\alpha$ probing \ion{H}{I} and \ion{Mg}{II} $\lambda 2796$.

The Ly$\alpha$ EW data are compiled using three samples: COS-Halos \citep{werk2013ApJS}, COS-GASS \citep{Borthakur2016ApJ}, and DIISC \citep{Borthakur2024arXiv}.
There are in total 111 galaxies included in the compiled sample.
Galaxies in these three samples exhibit similar stellar mass, peak at $10^{10.5}~\msun$, while the DIISC sample spans a relatively broader range.
Collectively, the 1-$\sigma$ scatter of the stellar mass is about $0.48$ dex for all 111 galaxies.
Whereas for the specific star formation rate (sSFR), the COS-Halos and the COS-GASS data present bimodal distribution including both star-forming and passive galaxies but the DIISC sample is dominated by blue galaxies.
Galaxies in the DIISC sample are more concentrated in the range of $-11 \lesssim \log {\rm sSFR/yr^{-1}} \lesssim -9.5$.
The combined sample of Ly$\alpha$ consists of $\sim 70$\% star-forming and $\sim 30$\% passive galaxies \footnote{Here, we follow the criteria in \citet{Borthakur2024arXiv}, which labels galaxies with $ \log {\rm sSFR/yr^{-1}} < -11$ as passive and the rest as blue galaxies.}.

The \ion{Mg}{II} sample is adopted from the \emph{Magellan MagE \ion{Mg}{II} (M3) Halo Project} \citep{huang2021MNRAS}, with 211 isolated galaxies included in our analyses.
Similar to the Ly$\alpha$ sample, these galaxies covers dwarfs with $L_{\rm B} \lesssim 0.1 L_{\rm B,*}$ to massive galaxies $L_{\rm B} \gtrsim 1 L_{\rm B,*}$.
The \ion{Mg}{II} absorption features are searched in the intermediate resolution spectra ($R\approx 4000$) obtained using MagE on the Magellan Clay Telescope.
In total, there are 85 sightlines with detected \ion{Mg}{II} features and 126 with 2$\sigma$ upper limits on \ion{Mg}{II} $\lambda2796$ EWs.

\section{Analyses and results}
\label{sec:analysis}
In this section, we fit the cool cloud CGM model to the gas density and EW samples to extract the gas density profile and the cool gas spatial distribution.
In particular, a two-step fitting method is adopted here to separate the gas density profile and the distribution of cool clouds.
We first obtain the gas density profile by fitting the cool gas density sample.
Then, this gas density profile is fixed in a joint fitting of the \ion{H}{I} Ly$\alpha$ and \ion{Mg}{II} $\lambda 2796$ EWs to obtain the spatial distribution of cool gas clouds.
In this section, we present the fitting methods and results.

\subsection{The fitting method}
\label{sec:fit_method}
In the following analyses, all fittings are performed under a Bayesian framework. 
First, we calculate the predicted PDFs of densities or EWs at different radii for a set of model parameters ($\theta$).
Then, the likelihood $\mathcal{L}_i ({\rm EW}_i | \theta)$ is calculated for individual measurements or limits of density or EW, which are independent from each other.
Finally, the total likelihood for the dataset of $n$ measurements and $m$ upper limits are

\begin{equation}
  \log \mathcal{L}(D|\theta) = \sum_i^n \log \mathcal{L}_i ({\rm EW}_i, \sigma_i | \theta) + \sum_i^m \log \mathcal{L}_i ({\rm EW}_{i, u} | \theta),
\end{equation}
where ${\rm EW}_i$ and $\sigma_i$ are the median and associated uncertainties for a measurement, while ${\rm EW}_{i, u}$ is a 2-sigma upper limit for non-detection.

The forms of $\mathcal{L}_i ({\rm EW}_i, \sigma_i | \theta)$ and $\mathcal{L}_i ({\rm EW}_{i, u} | \theta)$ are given by convolving the measured value with the model-predicted PDF of densities or EWs at their projected distance $d_{\mathrm{proj}, i}$.
In practice, the probability distribution of measured densities or EWs are assumed to be a normalized top-hat function within the 1$\sigma$ uncertainty as
\begin{equation}
    \mathcal{L}_i ({\rm EW}_i, \sigma_i | \theta) = \frac{{\rm CDF}({\rm EW}_i+\sigma_{i, u}) - {\rm CDF}({\rm EW}_i-\sigma_{i,l})}{\sigma_{i, u}+\sigma_{i, l}},
\end{equation}
where $\sigma_{i, u}$ and $\sigma_{i, l}$ are the upper and lower uncertainties, and CDF is the cumulative distribution function calculated from the model-prediction PDF.
This approximation of the likelihood is adopted for simplicity because of the significantly smaller uncertainty in observation ($\approx 1/10$) than the width of the modeled PDFs, considering the projection effect, intrinsic scatter, and cloud clumpiness.

\begin{figure*}
  \includegraphics[width=0.97\textwidth]{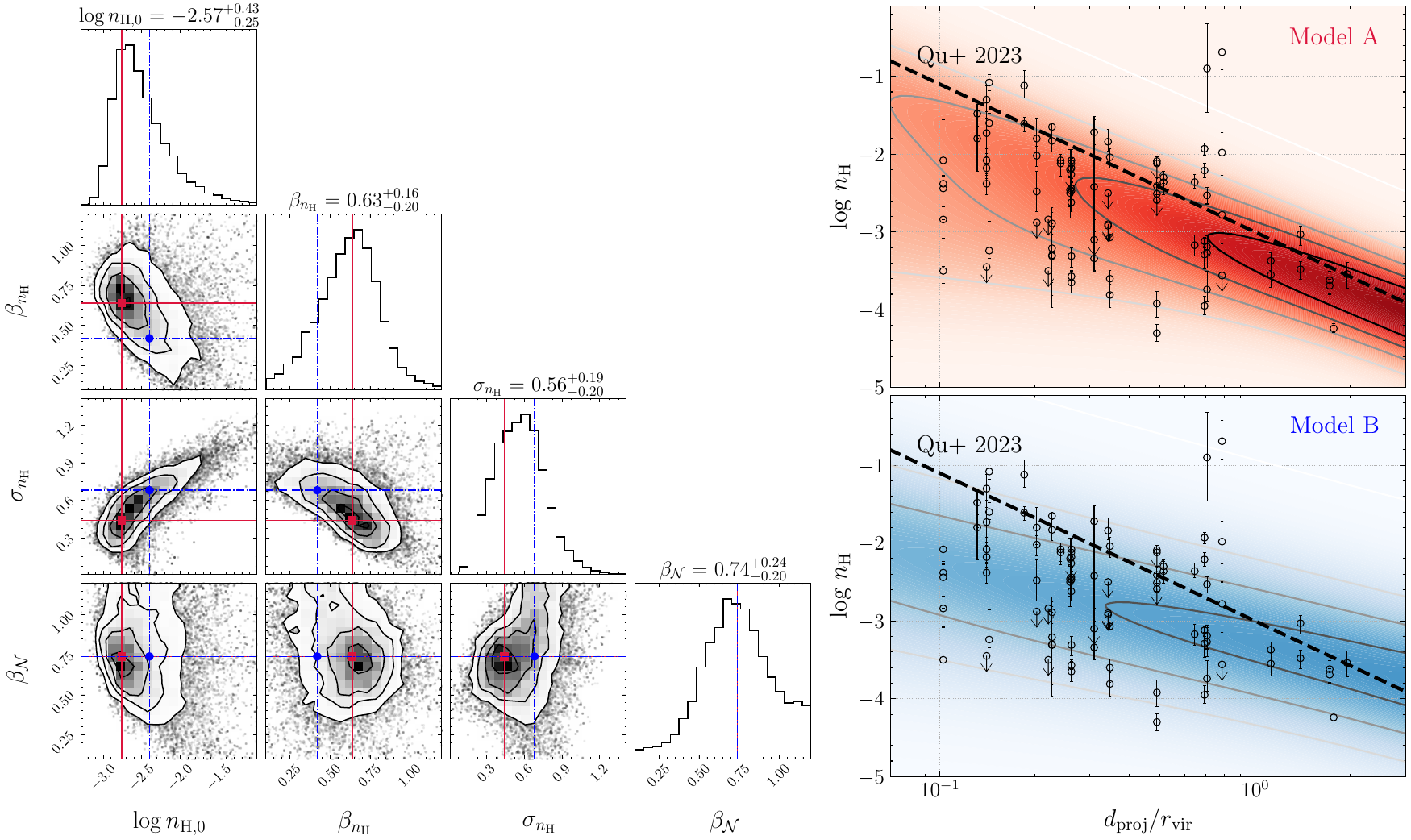}
  \caption{The best-fit model of the cool gas density $\nc$ sample.
  Left panel: the corner plot obtained in the MCMC approach.
  Two potential models are marked to represent two scenarios with different contributions to the gas density due to the projection effect and the intrinsic scatter.
  The red square is the maximal likelihood (Model $\nc$ A), while the blue circle is another solution within $1 \sigma$ away from the maximum likelihood (Model $\nc$ B), which exhibits a shallower gas density slope and a larger intrinsic gas density.
  Right panels: the 2D PDF of the gas density at different projected distances.
  The gas density scatter in Model $\nc$ (A) is from both projected and intrinsic scatter, whose radial profile of the gas density is consistent with the linear fit of the maximum density along each sight line in \citet[][]{qu2023MNRAS}.
  The Model $\nc$ (B) is dominated by the intrinsic scatter.
  Limited by the small sample, these two solutions cannot be distinguished.
  }
  \label{fig:nH_fit}
\end{figure*}

For upper limits, the likelihood is calculated with two terms as 
\begin{equation}
\label{eq:likelihood:upper}
    \mathcal{L}_i ({\rm EW}_{i, u} | \theta) = \epsilon(d_{\mathrm{proj}, i}) \frac{{\rm CDF}({\rm EW}_{i, u})}{ {\rm EW}_{i, u}} + [1-\epsilon(d_{\mathrm{proj}, i})],
\end{equation}
where $\epsilon=1-{\rm Pois}(0|\Ncl)$.
Here, the first term represents the probability of sightlines with low EWs below the detection limit,  which is calculated by integrating from 0 to the upper limit.
The second term represents the probability of hitting no clouds along the sightline (also see \citealt{huang2021MNRAS, Qu2024}).

We assume uniform priors and construct the posteriors of individual model parameters to reproduce the observed cool gas densities and the samples of Ly$\alpha$ and \ion{Mg}{II} $\lambda 2796$ EWs. %\qzj{may need revise}.
The best fit of these parameters is obtained using the Markov-Chain Monte-Carlo (MCMC) method, which is carried out with the Python module \texttt{emcee} \citep{foreman2013PASP}.
In the following analysis, the MCMC step is set to 2000, ensuring convergence with the Gelman-Rubin diagnostic \citep{Gelman1992}.

\subsection{Gas density profile}
\label{sec:nH_fit}

The first step in the two-step analysis is constraining the cool gas density profile using the gas density sample compiled in \citet{qu2023MNRAS}.
As stated in Section \ref{sec:density_dis}, the projection effect and the local variation of gas density lead to distinct features in the PDF of density as a function of projection radius.
Adopting the Bayesian framework described in Section \ref{sec:fit_method}, we in particular obtain the best-fit of $\normc$, $\betac$, $\sigma_{\nc}$, and $\betaN$, and the cloud mass is fixed to the fiducial value of $10^4 ~\msun$.
Different from the following EW fitting, here we fit the normalized PDF of gas density instead of the incidence rate.
This is because the gas density can only be obtained for metal-enriched clouds, whose incidence rate also depends on the sensitivity of QSO spectroscopy and the metallicity variation in the CGM, which are beyond the scope of this work.
Therefore, the normalization of $n_{{\scriptscriptstyle\mathcal{N}}_\mathrm{cl},0}$ cannot be determined using the gas density sample.
$\betaN$ is included because it affects the distribution of clouds at different radii, hence the detection probability along the line of sight.

\begin{table*}
	\centering
	\caption{Best-fit parameters of the $\nc$ and EW samples.}
	\label{tab:best_fit}
	\begin{tabular}{lccccccccccc}
\hline
\hline
& $\log n_{\rm H,0}$ & $\betac$ & $\sigma_{\nc}$ & $\log \normN$ & $\betaN$ & $\sub{b}{peak}$ & $\sigma_{\mathrm{left}}$ & $\sigma_{\mathrm{right}}$ & $\mcl$ & Predicted $\log M_{\rm cool}$\\
Model & $/\mathrm{cm}^{-3}$ & & dex & $/\rvir^{-3}$ & & \kms & \kms & \kms & $\msun$ & $/\msun$ \\
(1) & (2) & (3) & (4) & (5) & (6) & (7) & (8) & (9) & (10) & (11)\\
\hline
$n_{\rm H}$ & $-2.57^{+0.43}_{-0.25}$ & $0.63^{+0.16}_{-0.20}$ & $0.56^{+0.19}_{-0.20}$ & $...$  & $0.74_{-0.20}^{+0.24}$ & $...$ & $...$ & $...$ & $10^4$ & $...$ \\
[1ex]
$n_{\rm H}$ (A) & $-2.76$ & $0.64$ & $0.44$ & $...$ & $0.74$ & $...$ & $...$ & $...$ & $10^4$ & $...$ \\
[1ex]
$n_{\rm H}$ (B) & $-2.4$ & $0.42$ & $0.68$ & $...$ & $0.74$ & $...$ & $...$ & $...$ & $10^4$ & $...$ \\
\hline
EW & $-2.76$ & $0.64$ & $0.44$ & $4.76^{+0.27}_{-0.31}$ & $0.65^{+0.06}_{-0.07}$  & $25^{+17}_{-12}$ & $>20$ & $24^{+7}_{-12}$ & $10^4$ & $10.01^{+0.06}_{-0.06}$ \\
[1ex]
EW (mc3) & $-2.76$ & $0.64$ & $0.44$ & $6.56^{+0.28}_{-0.32}$ & $0.61^{+0.05}_{-0.05}$  & $9^{+10}_{-3}$ & $>20$ & $19^{+3}_{-4}$ & $10^3$ & $ 9.62^{+0.05}_{-0.07}$ \\
[1ex]
EW (mc6) & $-2.76$ & $0.64$ & $0.44$ & $3.50^{+0.32}_{-0.30}$ & $0.56^{+0.07}_{-0.06}$ & $24^{+13}_{-12}$ & $>20$ & $16^{+8}_{-12}$ & $10^6$ & $10.46^{+0.05}_{-0.05}$ \\
\hline
\multicolumn{11}{l}{$^{1}$ Model labels.}\\
\multicolumn{11}{l}{$^{2-4}$ Parameters of the cool gas density profile. Columns are the normalization value at $\rvir$, the slope, and the intrinsic scatter, respectively.}\\
\multicolumn{11}{l}{$^{5-6}$ Parameters of the cloud number density profile. Columns are the normalization value at $\rvir$ and the slope, respectively.}\\
\multicolumn{11}{l}{$^{7-9}$ Parameters of the split Gaussian distribution, used to describe the velocity dispersion of clouds (see Section \ref{sec:prob:EW}). Columns are peak value, the} \\
\multicolumn{11}{l}{left-side width, and the right-side width, respectively.}\\
\multicolumn{11}{l}{$^{10}$ Mass per cloud used in the corresponding model.}\\
\multicolumn{11}{l}{$^{11}$ Predicted total mass of the cool CGM according to the best-fit parameters.}\\
	\end{tabular}
\end{table*}

The posterior distributions of best-fit parameters are shown in Figure \ref{fig:nH_fit}, while the median and $1\sigma$ uncertainty are reported in Table \ref{tab:best_fit}.
The best-fit density at $\rvir$ is $\log \normc/\cc = -2.57^{+0.43}_{-0.25}$, and the slope of the cool gas density is $\betac = 0.63^{+0.16}_{-0.20}$.
Here, we note that the fitted $\nc$ at $\rvir$ is significantly higher than the observed gas density of $\log \nc/{\rm cm^{-2}} \approx -3$ at $\rvir$.
This is mainly because $\normc$ is the maximum density projected at $\rvir$, while both the projection effect and the cross-section correction lead to more detectable low-density clouds, which eventually results in a lower median of the $\nc$ PDF at $\rvir$ (see Section \ref{sec:density_dis} and Figure \ref{fig:nH_comparison}). 

In addition to the density radial profile, the local density variation exhibits a $\sigmac = 0.56_{-0.20}^{+0.19}$ dex in the logarithmic scale.
Such an intrinsic scatter suggests that the cospatial cool gas with different densities may not be in the pressure equilibrium between each other, considering the relatively constant PIE temperature dependence on the density.
This expectation of pressure imbalance is consistent with the direct measurements of the multiphase component, showing a potential pressure difference of a factor of $\approx 10$ \citep{qu2022MNRAS}.

Here, we consider whether the observed scatter in observations is dominated by local density variation or projection effect.
In particular, two potential models are within the $\approx 1\sigma$ from the median of the posterior distributions of all parameters (i.e., Models $n_{\rm H}$ A and B in Table \ref{tab:best_fit}).
The Model $n_{\rm H}$ (A) has $\log \normc/\cc = -2.76$, $\betac = 0.64$, and $\sigmac = 0.44$, which is about the maximum likelihood solution.
The Model $n_{\rm H}$ (B) is $\log \normc/\cc = -2.40$, $\betac = 0.42$, and $\sigmac = 0.68$, which is $\approx 1\sigma$ away from the maximum likelihood solution.
As discussed in Section \ref{sec:density_dis}, the Model $n_{\rm H}$ (A) prefers a more significant contribution from the projection effect into the observed density scatter.
However, Models $n_{\rm H}$ (A) and (B) cannot be distinguished using the current gas density sample.
The predicted PDFs of the gas density of Models $n_{\rm H}$ (A) and (B) are plotted in the right panels of Figure \ref{fig:nH_fit}.
We also compare the relation obtained by assuming the highest gas density along each sight line occurs at $\rproj$, which is $\log (n_{\rm H}/\cc) = (-1.9\pm 0.3) \times \log (d_{\rm proj}/r_{\rm virial})] + (-3.0\pm0.1)$.
The best-fit cool cloud Model $\nc$ (A) exhibits consistent parameters, verifying the adopted assumption in \citet{qu2023MNRAS}.

The most obvious difference between the two scenarios is the slope of the gas density profile.
In the projection scenario (Model $\nc$ A), a larger power-law slope of $\betac \approx 0.64$ is required to have a significant variation of 2 dex in the gas density from the inner halo to the outskirts ($0.1$--$1 \rvir$).
Then, the observed density scatter of $2-3$ dex at small projected distances is dominated by the projection effect.
In the intrinsic variation scenario (Model $\nc$ B), a power-law slope of $0.4$ leads to a density difference of only $\approx 1$ dex from $0.1$--$1 \rvir$.
Then, the observational scatter at small projected distances is less contributed by the projection effect, instead, a larger intrinsic scatter is needed (i.e., 0.68 dex compared to $0.44$ dex in the projection scenario).

The major difference between these two scenarios is the widths of the cool gas density distribution at different radii, as presented in Figure \ref{fig:nH_comparison}.
The projection scenario prefers a wider distribution at smaller projected distances, while the scatter scenario exhibits similar widths over various projected distances.
In the observations, it is clear that inner sightlines exhibit a larger scatter of 2--3 dex in the gas density, while the scatter in the outskirts is smaller, also see \citet{Sameer2024} for similar trends with different assumptions on the ionization mechanism.

\subsection{Joint fit of \Lya\ and \ion{Mg}{II}}
\label{sec:joint_lya_mgII}

With the measured density profile, we further constrain the number density profile of cool clouds and the velocity dispersion of absorption features using a joint fit of the EWs of Ly$\alpha$ and \ion{Mg}{II} $\lambda 2796$.
In particular, the column densities of different species are calculated using PIE models with the assumed cloud mass of $10^4\,\msun$, and are converted to EWs with the distribution of Doppler $b$ factor characterized with three free parameters (see details in Section \ref{sec:model}).
Despite the distinct systematic redshifts of \ion{Mg}{II} and \Lya\ samples, the effect of UVB variation is minor (discussed further in Section \ref{sec:caveats}).
Therefore, we only use the incident field produced by the HM05 field at $z=0.22$ (median value of the \ion{Mg}{II} sample).

In this joint fit of Ly$\alpha$ and \ion{Mg}{II} $\lambda 2796$ EW samples, we adopt the maximum likelihood parameters of the density profile fitting $\normc=-2.76$, $\betac=0.64$, and $\sigmac=0.44$, i.e., the maximum likelihood solution reported in Section \ref{sec:nH_fit} and Table \ref{tab:best_fit}.
The parameter $\betaN$ is still a free parameter, because it is sensitive to the absorption strength, leading to strong constraints in the EW fit rather than the $\nc$ fit.
For the velocity dispersion, we adopt a uniform prior of $10$ to $200$~\kms, where the lower boundary is about the minimal of Doppler $b$ of individual absorption components \citep[e.g.,][]{werk2013ApJS, qu2023MNRAS}, while the upper bound is about the maximum of light-of-sight velocity dispersion \citep[e.g.,][]{Qu2024}.

\begin{figure*}
  \includegraphics[width=0.8\textwidth]{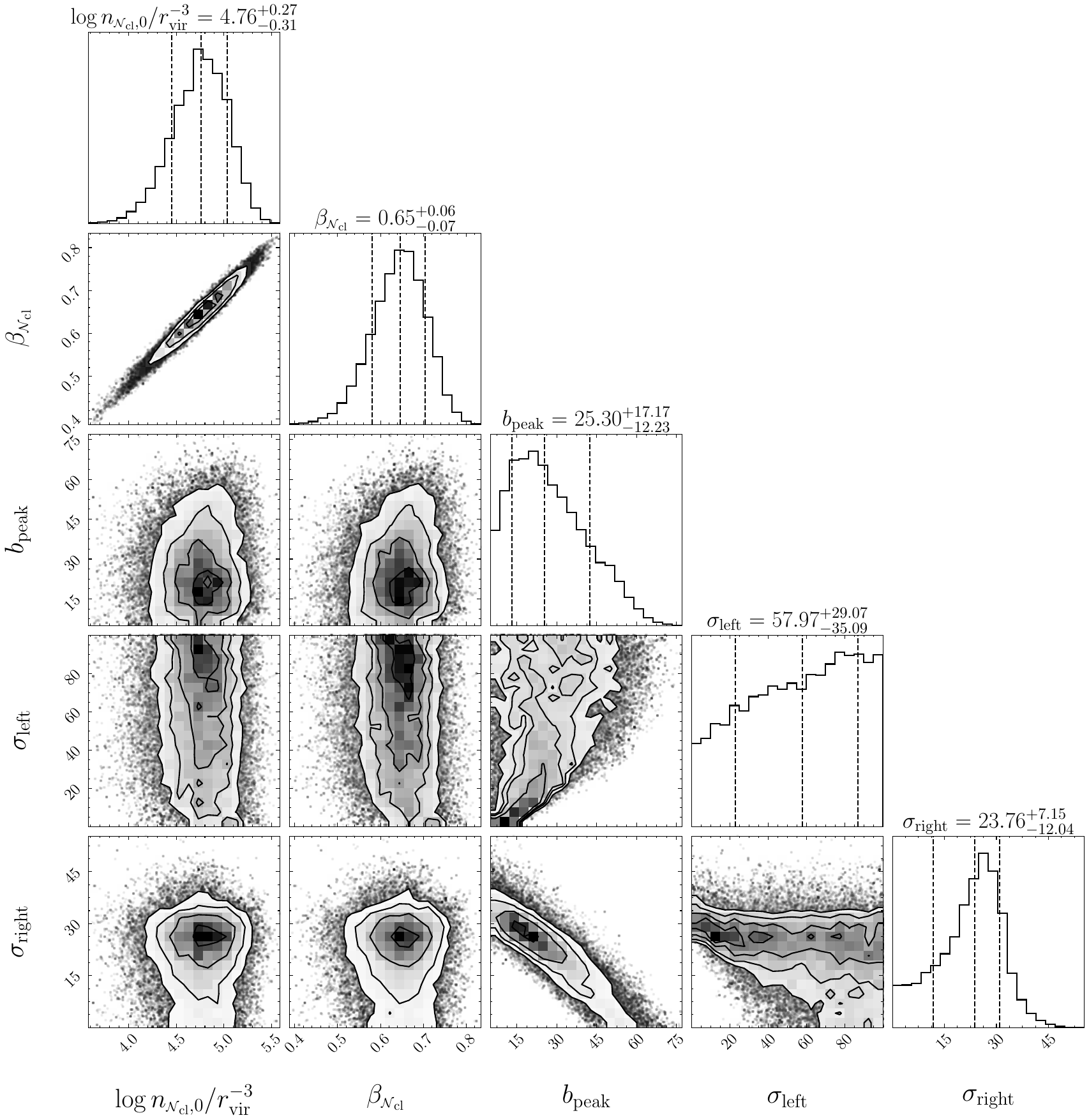}
  \caption{The corner plot for the joint fitting of Ly$\alpha$ and \ion{Mg}{II} $\lambda 2796$ EW samples. In this fitting, the density profile is fixed to the maximum likelihood model shown in Figure \ref{fig:nH_fit} (i.e., the Model ``$n_{\rm H}$(A)'' in Table \ref{tab:best_fit}).
  }
  \label{fig:corner_ew}
\end{figure*}

The best-fit parameters of the CGM model for the EW samples are reported in Table\ref{tab:best_fit} (i.e., Model ``EW''), and the corner plot is shown in Figure \ref{fig:corner_ew}.
With the fixed density profile, the cloud number density profile exhibits parameters of $\log \normN / \mathrm{kpc}^{-3} = -2.12^{+0.27}_{-0.21}$ and $\betaN = 0.65^{+0.06}_{-0.07}$.
Here, $\betaN$ is also constrained when fitting the gas density sample.
The EW-fit $\betaN$ is consistent with the $\nc$ fit within $\approx 0.5 \sigma$. 
However, it exhibits a smaller uncertainty of 0.05 than the $\nc$ fit, because the number of clouds is more sensitive to the absorption strength rather than the PDF contained in the EW fit.
Such a profile leads to about $\approx3$ clouds along the sightline projected at the virial radius, and $\approx10$ clouds at the inner halo.
Based on the best-fit number of clouds, we also predict the cool CGM mass, which will be further discussed in Section \ref{sec:discuss:mcl}.

The LOS velocity dispersion is also constrained in the EW fit because of the conversion from the modeled column density to EW.
In the cool CGM model, we adopt a split Gaussian distribution of the LOS velocity at different radii, defined in Section \ref{sec:prob:EW}.
In particular, the peak LOS velocity is about $\approx 20-30$ \kms, which is consistent with observed and simulation-predicted LOS velocities of metals \citep[e.g.,][]{Koplitz2023, Qu2024}.
The left-side Gaussian width is poorly constrained as a $2\sigma$ lower limit of $\approx 20$ \kms, leading to a relatively flat distribution down to the lower boundary of $10$ \kms.
Because the lower Doppler $b$ values cannot be ruled out at $3 \sigma$, the best fit is insensitive to the lower boundary of allowed Doppler $b$ values.
On the right side, the tail of the velocity dispersion distribution exhibits a Gaussian $\sigma$ of $22$~\kms.
This is also consistent with observations of LOS velocities, where only sightlines projected the inner halo ($\lesssim 0.5\rvir$) exhibit broad features with $\vdisp>60$ \kms~ (i.e., $2\sigma$ in the best-fit model).

\begin{figure*}
\centering
  \includegraphics[width=0.9\textwidth]{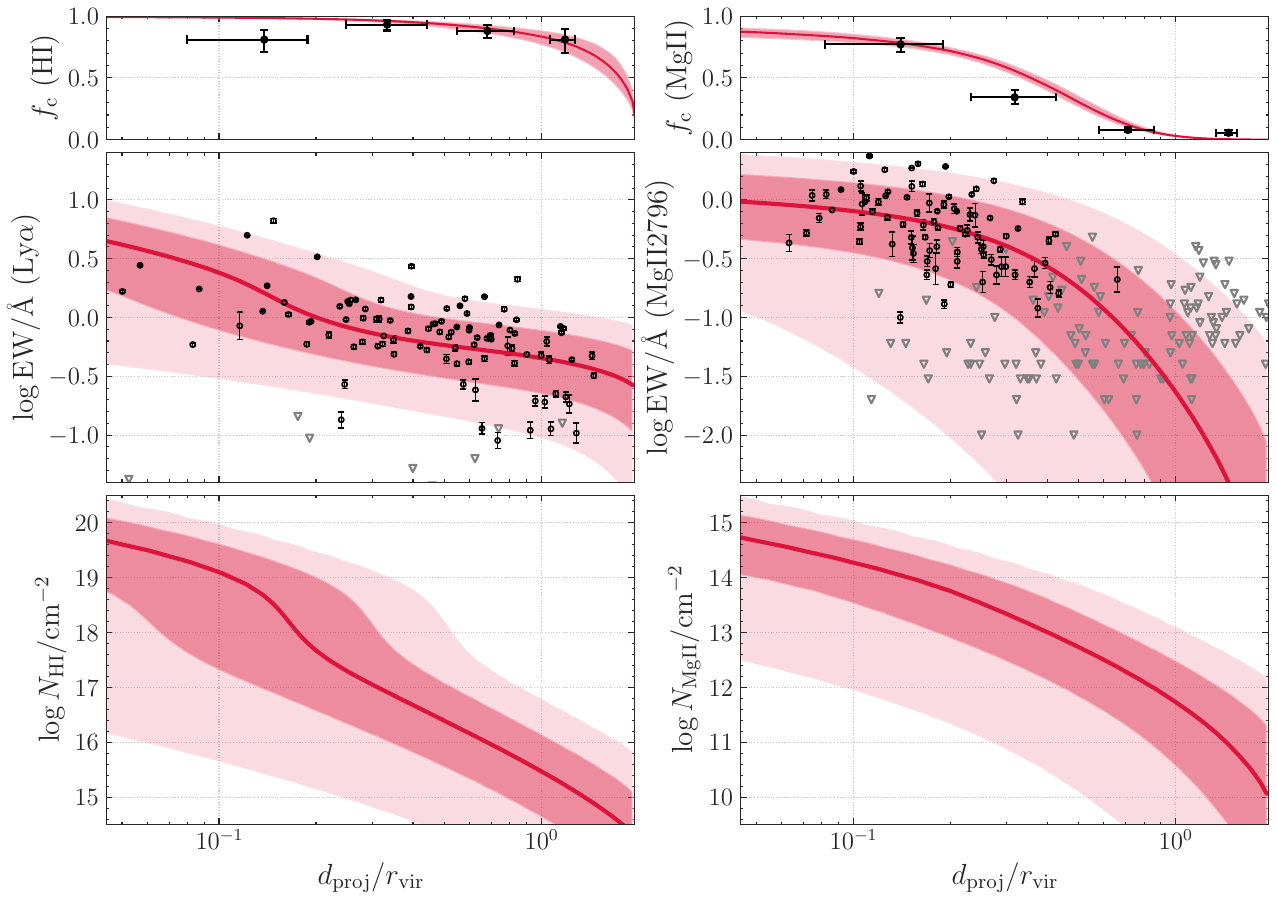}
  \caption{The best-fit models of the joint EW samples of Ly$\alpha$ and \ion{Mg}{II} $\lambda2796$ in the left and right panels, respectively.
  Top row: the covering fraction of the observed sample (circles) and the model prediction (shaded regions).
  The covering fractions are calculated with limiting EWs of $\log {\rm EW}/$\r{A} $=-0.8$ and $-0.5$ for Ly$\alpha$ and \ion{Mg}{II} $\lambda 2796$, respectively.
  Middle row: the comparison of the EWs and the models.
  Detected absorption features are shown as open circles with error bars, while upper limits are downward triangles.
  The darker and lighter shaded regions correspond to the 1- and 2-$\sigma$ errors, respectively.
  Bottom row: the predicted column density.}
  \label{fig:joint_fit}
\end{figure*}

The best-fit PDF of the EW is compared with observations in Figure \ref{fig:joint_fit} for both \ion{H}{I} and \ion{Mg}{II}.
%\qzj{introduce the observed EWs: relatively narrow dynamic range for both HI and MgII.}
Overall, the best-fit model reproduces the observed EW distribution, as seen in the middle row in Figure \ref{fig:joint_fit}.
For \ion{H}{I}, the observed EWs exhibit relatively flat radial profiles, with the median value declining about 1 dex from inner ($0.1 \rvir$) to outer ($\rvir$) sightlines. 
As a comparison, the \ion{H}{I} column density typically exhibits a larger decline (e.g., \citealt{Chen2001}).
The difference between the column density and the EW is mainly due to saturation, and the observed scatter in EWs is significantly affected by the velocity dispersion.
In the bottom panels, we also present the distributions of \ion{H}{I} and \ion{Mg}{II} column densities.
The median column density of \ion{H}{I} decreases from $\sim 10^{19}$ to $\sim 10^{15}$ \pcm~ from the inner halo to the outskirts.
For \ion{Mg}{II}, the absorption features are only detected within $\approx 0.5 \rvir$, while there are only non-detections beyond  $\rvir$.
The \ion{Mg}{II} column density decreases by $2$ dex from 0.1 to $1\rvir$.

Next, we compare the covering fraction of Ly$\alpha$ and \ion{Mg}{II} $\lambda 2796$ with the best-fit model.
In particular, the limiting EWs of \Lya\ and \ion{Mg}{II} $\lambda$2796 are set to $\log\,$EW/\r{A} $= -0.8$ and $-0.5$, respectively for calculation of the covering fraction.
As shown in Figure \ref{fig:joint_fit}, the observed covering fractions are reproduced by the best-fit model.

\section{Discussion}
\label{sec:dis}
Applying this cool cloud CGM model to the observation samples, we constrain the gas density profile and number density of cool clouds.
Here, we discuss the caveat in this model and the implications on the density fluctuation in the cool CGM and the total cool CGM mass.

\subsection{Caveats in the cool cloud CGM model}
\label{sec:caveats}

As described in Section \ref{sec:model}, multiple assumptions are adopted to simplify the cool cloud CGM model, including (1) the uniform profiles and distributions in halos with different masses, (2) the incident radiation field for photo-ionization, and (3) the same distribution of velocity dispersion along at different radii.
Here, we discuss the potential caveats due to these assumptions. 

First, limited by the rarity of background QSOs, most galaxies in the adopted sample only have one QSO sightline probing the cool CGM.
Then, the cool CGM model is applied to a sample of galaxies with different masses and star formation rates.
As introduced in Section \ref{sec:data}, the adopted EW samples are mainly composed of star-forming galaxies with $\mstar\approx 10^{10}-10^{11} \msun$.
Therefore, there are potential halo-by-halo variations of the cool CGM distribution. 
In this work, we adopt an assumption that these galaxies share similar cool gas properties, after normalizing the radius with $\rvir$.
This assumption is commonly adopted in absorption studies, while there are also studies suggesting that the absorption strength depends on stellar mass \citep[e.g.,][]{Chen2001, Bordoloi2018}.
Also, recently deep \ion{H}{I} 21 cm observations reveal that \ion{H}{I} column densities may be self-regulated down to $N\approx 10^{18} {\rm~ cm^{-2}}$, as well aligned \ion{H}{I} column density profile using the radius at $\Sigma_{\rm HI} = 0.01 \msun~{\rm pc^{-2}}$ (i.e., $\approx 1.2\times 10^{18}\rm~ cm^{-2}$), rather than following the $\rvir$ normalization \citep{Borthakur2024arXiv, Wang2025}.
It will be further investigated in the following studies on the connection between the absorption-detected cool CGM and the \ion{H}{I} 21 cm line-detected ISM-CGM interface.

Then, in the fiducial model, we adopt the incident radiation field as HM05 UVB at $z = 0.22$ in the cloudy models, calculating column densities of \ion{H}{I} and \ion{Mg}{II}.
Systematic uncertainties can be introduced due to different UVB templates, redshifts, and potential escaping ionizing photons from the host or nearby galaxies.
To examine the impact of the varied incident field on the model, we consider UVB models from \citet{Haardt2001, Khaire2019, FG20}, and include the escaping flux from the host galaxy.
In particular, the luminosity escaping flux is estimated based on the availability of SFR of $L^*$ galaxies \citep[e.g.,][]{Renzini2015}, the SFR-UV flux conversion of \citet{Fumagalli2012ApJ, Bouwens2016ApJ}, and assumed escaping fraction of 0.1.
Because the escaping flux declines with the radius as $r^{-2}$, the escaping flux only dominates $\lesssim  0.3\rvir$ at low redshift.

We test a series of incident fields with different combinations of UVB templates and redshifts. 
Varying the UVB models plays a minor role in regulating the ion fractions.
This is because the ionization parameter $\log U$ determines the ionization state,  and varying the UVB intensity will lead to different estimations of gas densities, rather than affect the ionization fractions for \ion{H}{I} and \ion{Mg}{II}.
In this work, we adopt the HM05 UVB, consistent with the gas density sample compiled in \citet{qu2023MNRAS}.

\begin{figure}
    \centering
    \includegraphics[width=0.97\columnwidth]{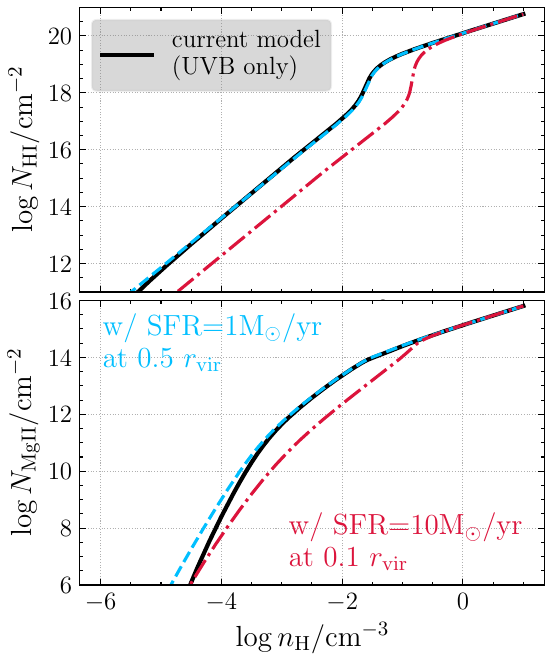}
    \caption{The Cloudy models with different assumed radiation fields. The black curves show the fiducial model without galactic radiation. The red and blue curves are two cases with strong and weak escaping fluxes from the host galaxies, which is discussed in Section \ref{sec:caveats}.
    }
    \label{fig:UVB_variation}
\end{figure}

More significant differences are seen in models with escaping ionizing photons from host galaxies.
Here, we use two examples in Figure \ref{fig:UVB_variation} to show the differences in the ionization fractions of \ion{H}{I} and \ion{Mg}{II} induced by the escaping photons.
Compared to the fiducial model, the low escaping flux case (0.5 $\rvir$ away from low SFR galaxies) leads to a negligible difference over different densities as expected.
The high escaping flux (a high SFR and a small $\rproj = 0.1 \rvir$) leads to reduced $N_{\rm HI}$ and $N_{\rm MgII}$ by $\approx 1$ dex when the gas density is about $\log \nc/{\rm cm^{-2}} \lesssim -1$. 
Therefore, gas phases may be much different at the interface between ISM and CGM for starburst galaxies, which is beyond the scope of this work.

In addition to the incident radiation field, another major assumption in this model is about the velocity dispersion, which dominates the conversion from column densities to EWs, when the absorption is saturated.
The velocity dispersion has contributions from both the internal velocity broadening within clouds and the bulk velocity difference along the sight line \citep[e.g.,][]{Lopez2024}.
The observational constraints on the velocity dispersion of \ion{H}{I} and \ion{Mg}{II} are still limited.
Using \ion{O}{VI} as a tracer, the velocity dispersion is found to have a narrow dynamic range of $15-60$ \kms~ at $\gtrsim 0.2 \rvir$ \citep{Qu2024}.
The \ion{H}{I} and \ion{Mg}{II}-traced phase is denser than the \ion{O}{VI}-bearing phase, leading to narrower features.
In this model, we assume the same distribution of the velocity dispersion for both \ion{H}{I} and \ion{Mg}{II} at all radii (Section \ref{sec:prob:EW}), and obtain a median $\vdisp \approx 20-30$ \kms~ and a tail of broad features up to $\approx 60$ \kms.
Imposing a declining $b_{\rm peak}$ over the radii may affect the obtained $\betaN$, because larger Doppler $b$ leads to higher EW values, hence lower column densities to reproduce the observed EWs. 
However, limited by the current sample, constraints on the velocity dispersion are infeasible.

\subsection{Cloud mass variation}
\label{sec:discuss:mcl}

As mentioned in Section \ref{sec:assump:mcl}, the cloud mass in the fiducial model is fixed at an empirical value of $10^4 ~\msun$.
But the cloud size-density relation \citep{chen2023ApJ, Sameer2024} also indicates a large dispersion of cloud mass.
Using a different cloud mass in the model not only changes the cloud sizes, but also alters the ionization level within clouds, hence producing different ion column densities for each cloud.
Furthermore, to reproduce the observed EWs and total column densities of ions, the number of clouds per sightline should be decided accordingly.
In combination, we may get a distinctive fitting result as well as its interpretation.
In this section, we discuss the potential impact of other cloud mass values.

According to the cloud size-density relation in \citet{chen2023ApJ} and \citet{Sameer2024}, the cloud mass range spans $3 \lesssim \log \mcl/\msun \lesssim 6$.
Therefore, we test our model with two other different mass scales of $\log \mcl/\msun = 3$ (model denoted as ``mc3'') and $\log \mcl/\msun = 6$ (model denoted as ``mc6'').
For these two runs, the fitting procedures follow that of the fiducial model, with the same gas density profile.
Here, we note that varying the cloud mass does not affect the gas density profile significantly, because $\mcl$ is degenerate with the number of clouds, which is unconstrained in the density fitting.
All best-fit parameters are within $0.5 \sigma$ from the ``mc4'' model for both ``mc3'' and ``mc6''.
The best-fit results are summarized in Table \ref{tab:best_fit}.

Varying the individual cloud mass does not dramatically change the slope of the number density profile of cool clouds.
The slopes remain roughly consistent within their confidence intervals for the three models.
On the other hand, it is expected that the absolute value of cloud number density reduces for higher cloud mass, and vice versa.
The number of clouds per volume space at the virial radius increases by $\sim 1.80$ dex in Model ``mc3'', and decreases by $\sim 1.26$ dex in Model ``mc6''.
The variation in the number density of cool clouds lead to different total cool CGM model, which will be further investigated in Section \ref{sec:mass}.

The velocity dispersion distribution is yet badly constrained, especially for the left-side width.
Interestingly, the peak value in model ``mc3'' shifts towards the lower end compared with the other two models. 
This suggests that clouds produce narrower absorptions in this model.

\begin{figure*}
    \centering
    \includegraphics[width=0.9\textwidth]{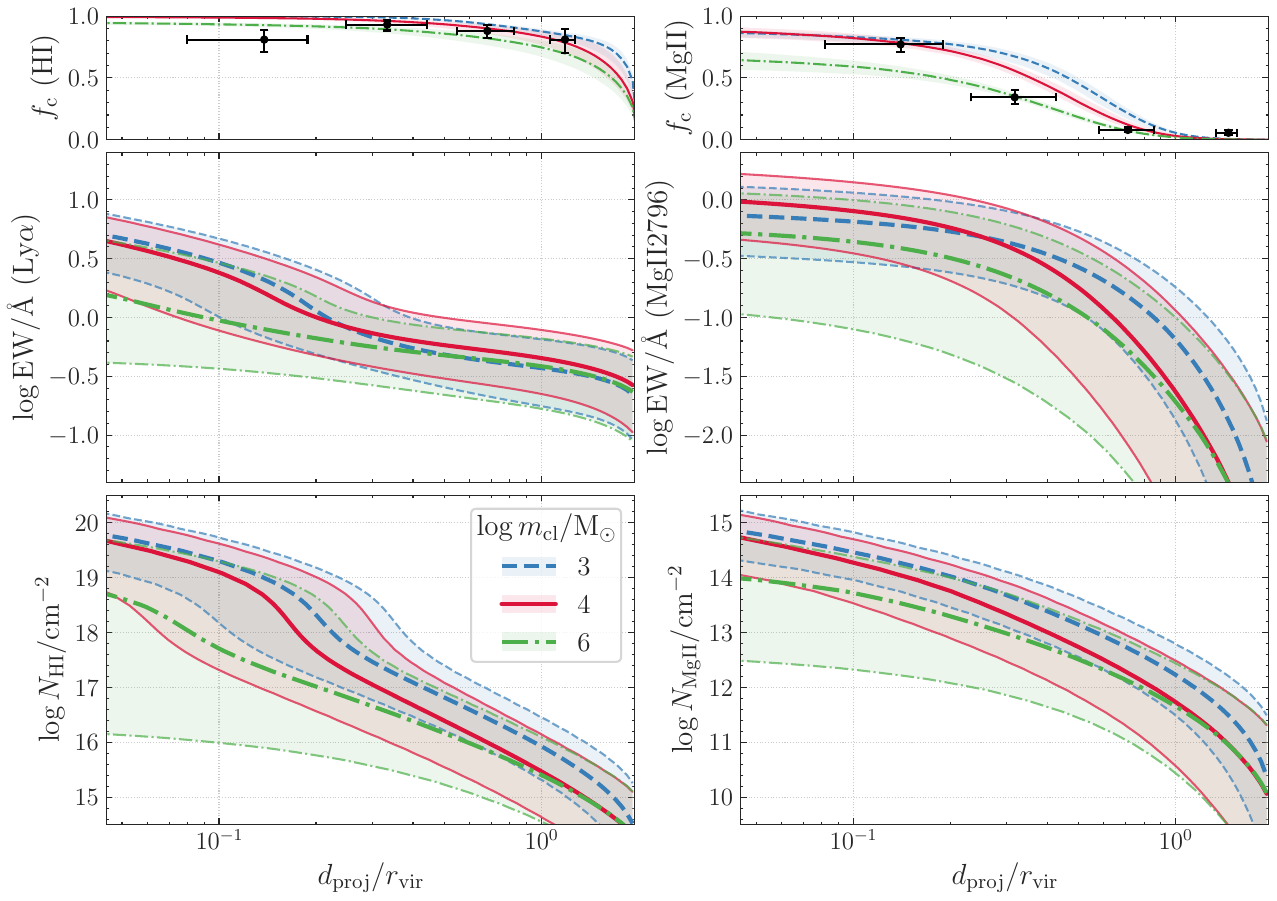}
    \caption{
    The best-fit results of models with different assumptions of individual cloud mass.
    The left and right columns are results for \Lya\ and for \ion{Mg}{II}, respectively.
    Top row: the covering fraction of the observed sample (solid circles) and the model prediction (shaded regions).
    The covering fractions are calculated with limiting EWs of $\log {\rm EW}/$\r{A} $=-0.8$ and $-0.5$ for Ly$\alpha$ and \ion{Mg}{II} $\lambda 2796$, respectively.
    Middle row: the predicted EWs with different cloud masses.
    Bottom row: the predicted column densities with different cloud masses.
    The shaded regions of each color represent the 1-$\sigma$ errors of the corresponding model.
    The medians and edges are shown as lines in different styles for each mass.
    Here, a cloud mass of $10^4~\msun$ is the fiducial model.}
    \label{fig:mclvar}
\end{figure*}

A comparison of physical quantities among the three models is shown in Figure \ref{fig:mclvar}.
It is hard to disentangle these models based on the EW and ion column distributions, except that model ``mc6'' exhibits a wider uncertainty.
Also, model ``mc6'' shows a significantly lower ($>3 \sigma$) covering fraction of $\approx 0.6$ for \ion{Mg}{II} than observations, reaching $\approx  0.8$ at impact parameter of $\rproj \approx 0.1-0.2 \rvir$.
This suggests that the ``mc6" model is less favored compared to lower cloud masses.
At the meantime, the $10^3~\msun$ model overestimates the covering fraction of \ion{Mg}{II} at $0.3-0.5 \rvir$.
Therefore, the fiducial model with an average cloud mass of $10^4~\msun$ is preferred.

\subsection{Total mass of the cool CGM}
\label{sec:mass}
The mass budget is a fundamental question as the missing baryon problem in galaxies \citep[e.g.,][]{McGaugh2010}.
Among different phases in the CGM, the cool CGM mass has been controversial in the past decade.
By modeling the photoionized column densities in the COS-Halos sample, \citet{Werk2014} reported a cool CGM mass of $\approx 10^{11}~ \msun$ around $L^*$ galaxies (also see \citealt{prochaska2017ApJ}).
However, this cool CGM mass might be overestimated due to several high EW/$N$ systems in the COS-Halos sample.
\citet{Bregman2018ApJ} suggested that these high column systems may be outliers, encountering extended gaseous disk, leading to $\approx 5$ times higher total mass assuming spherical geometry.
Excluding these high-column outliers, \citet{faerman2023ApJ} found a cool CGM mass of $\approx 10^{9}-10^{10}~\msun$ using the COS-Halos sample.

Another way to deal with these potential outliers is constructing full PDFs at all radii, which is the key point of the cool cloud CGM model presented here.
With the full PDFs of the column density or EW at different radii, the effect of these outliers is reduced.
Therefore, we include all available data points in the modeling, and examine the derived cool CGM mass from the model.
Based on the best-fit parameters, the total mass of the cool phase gas in CGM of $L*$ galaxies (i.e., the Ly$\alpha$ and \ion{Mg}{II} samples at $z\approx 0.1-0.4$) can be calculated by:
\begin{equation}
  M_{\rm cool} = \mathcal{N}_\mathrm{cl, tot}\,\mcl = \int_0^{\rvir} \nN 4\pi r^2 \diff r\,\mcl.
\end{equation}
For the fiducial model with $\mcl = 10^4~ \msun$, the total mass of cool CGM is $\log M_{\rm cool}/\msun = 10.01^{+0.06}_{-0.06}$, which is consistent with \citet{Faerman2024}.
For the other two models with different cloud masses, the predicted masses are $\log M_{\rm cool}/\msun = 9.62^{+0.05}_{-0.07}$ and $\log M_{\rm cool}/\msun = 10.46^{+0.05}_{-0.05}$ for cloud masses of $\mcl = 10^3 ~\msun$ and $10^6~ \msun$, respectively.
Therefore, the derived total cool CGM mass is about $\log M_{\rm cool}/\msun = 10.0$ dex for $L^*$ galaxies, with a systematic uncertainty of $0.4$ dex.

\section{Conclusions}
\label{sec:sum}
In this paper, we introduce a cool CGM model to account for the clumpiness in the cool gas.
This model is applied to the measurements, accounting for the PDF of gas density or observed absorption strength.
Here, we summarize the key findings:

\begin{itemize}
    \item The dispersion seen in the observed PDF of $\nc$ is contributed by both the projection effect and the intrinsic dispersion of the gas density.
    The best-fit model suggests a global decline of the cool gas density with a power-law slope of $3 \betac \approx 2$, together with an intrinsic dispersion of $\approx 0.5$ dex (i.e., Model $\nc$ A in Table \ref{tab:best_fit}).
    However, limited by the current sample size, we cannot rule out another possibility with more contribution from intrinsic dispersion (the Model $\nc$ B).
    \item With the obtained density profile, we fit the number density of clouds by applying this model to the EW sample.
    We found a significant decline in the number of clouds from the inner halo to the outskirts.
    The radial profile of the density of the cloud number follows a power law with a slope of $3\betaN\approx 2$.
    With this cool gas distribution, the total cool CGM mass is $\log M_{\rm cool}/\msun = 10.01^{+0.06}_{-0.06}$.
    \item We also examine the impact of the assumed cloud mass in this model. 
    By varying the mass per cloud from $10^3 ~\msun$ to $10^6 ~\msun$, we found that the main results remain somewhat similar, such as the roughly consistent slopes of $\betaN$.
    The total mass of the cool CGM varies from $\log M_{\rm cool}/\msun = 9.62_{\rm 0.07}^{+0.05}$ to $10.4 \pm 0.05$.
    As shown in Figure \ref{fig:mclvar}, the model with a cloud mass of $10^4~\msun$ is the preferred model.
\end{itemize}

\section*{Acknowledgements}
The authors thank the referee for the constructive suggestions, which improved the draft significantly.
The authors thank the University of Michigan for their support and for hosting the three authors, Yang, Qu, and Bregman, for this work.
Yang acknowledges the China Scholarship Council for sponsoring the study at the University of Michigan.
Ji acknowledges the science research grants from the China Manned Space Project and the National Natural Science Foundation of China (NFSC) Grant No. 12233005.
%%%%%%%%%%%%%%%%%%%%%%%%%%%%%%%%%%%%%%%%%%%%%%%%%%
\section*{Data Availability}
The data underlying this paper will be shared on reasonable request
to the corresponding author.



\bibliographystyle{mnras}
\bibliography{mnras} % if your bibtex file is called example.bib

%\begin{thebibliography}{99}
%\end{thebibliography}

%%%%%%%%%%%%%%%%%%%%%%%%%%%%%%%%%%%%%%%%%%%%%%%%%%

%%%%%%%%%%%%%%%%% APPENDICES %%%%%%%%%%%%%%%%%%%%%

%%%%%%%%%%%%%%%%%%%%%%%%%%%%%%%%%%%%%%%%%%%%%%%%%%

% Don't change these lines
\bsp	% typesetting comment
\label{lastpage}

\end{document}